\documentclass{article}
\usepackage{amssymb}
\usepackage{amsfonts}

\input{tcilatex}
\begin{document}

\title{An epistemic interpretation of quantum probability via contextuality}
\author{Claudio Garola \\
Department of Mathematics and Physics, University of Salento\\
Via Arnesano, 73100 Lecce, Italy\\
e-mail: garola@le.infn.it}
\maketitle

\begin{abstract}
According to a standard view, quantum mechanics (QM) is a contextual theory
and quantum probability does not satisfy Kolmogorov's axioms. We show, by
considering the macroscopic contexts associated with measurement procedures
and the microscopic contexts ($\mu $\textit{-contexts}) underlying them,
that one can interpret quantum probability as epistemic, despite its
non-Kolmogorovian structure. To attain this result we introduce a predicate
language $L(x)$, a classical probability measure on it and a family of
classical probability measures on sets of $\mu $\textit{-contexts}, each
element of the family corresponding to a (macroscopic) measurement
procedure. By using only Kolmogorovian probability measures we can thus
define \textit{mean conditional probabilities} on the set of properties of
any quantum system that admit an \textit{epistemic interpretation }but are
not bound to satisfy Kolmogorov's axioms. The generalized probability
measures associated with states in QM can then be seen as special cases of
these mean probabilities, which explains how they can be non-classical and
provides them with an epistemic interpretation. Moreover, the distinction
between compatible and incompatible properties is explained in a natural
way, and purely theoretical classical conditional probabilities coexist with
empirically testable quantum conditional probabilities.\medskip

\textbf{Key words}: quantum probability, contextuality, entanglement,
quantum measurements.
\end{abstract}

\section{Introduction}

There are some typical features of quantum mechanics (QM) that are well
established and accepted in the current literature but still raise
interpretative problems. We are especially interested here in the following
topics.

(i) Non-kolmogorovian character of quantum probability, implied by the
non-distributivity of the lattice of (physical) properties, which is the
basic structure of standard quantum logic (QL).

(ii) The doctrine that, whenever a physical system in a given state is
considered, a quantum observable generally has not a prefixed value but only
a set of \textit{potential} values, and that a measurement \textit{actualizes%
} one of these values, yielding an outcome that depends on the specific
measurement procedure that is adopted (\textit{contextuality}).

There is a huge literature on these topics, which goes back to the early
days of QM. We limit ourselves here to recall that the QL issue was started
by a famous paper by Birkhoff and von Neumann (1935), while the
contextuality at a distance (or \textit{nonlocality}) and, more generally,
the contextuality of QM were accepted by most physicists as "mathematically
proven" after the publication of Bell's (1964, 1966) and Kochen-Specker's
(1967) theorems, later supported by numerous different proofs of the same or
similar theorems (among which the famous proof of nonlocality provided in
1990 by Greenberger, Horne, Shimony and Zeilinger, which does not resort to
inequalities).

Non-classical probability and contextuality can be linked, and inquiring
their links leads to important achievements. This issue has been already
studied by the "V\={a}xj\"{o} school", and in particular by Khrennikov
(2009). We propose in this paper a new perspective, according to which
quantum probability and its nonclassical features can be interpreted as
derived notions in a classical probabilistic framework by taking into
account microscopic and macroscopic contexts.

To the best of our knowledge, our proposal is innovative. Let us therefore
summarize the essentials of it.

First of all, we introduce some epistemological and physical remarks on QM
in Section 2 by referring to a conception of QM according to which QM deals
with individual examples of physical systems (briefly, \textit{individual
objects}) and their properties (see, e.g., Busch et al., 1996). Bearing in
mind these remarks, we work out in Section 3 a predicate language $L(x)$
whose predicates either denote \textit{states} or pairs made up of a \textit{%
property} $E$ and a (generally unknown) \textit{microscopic context} ($\mu $%
\textit{-context}) $C$. Hence the elementary sentences of $L(x)$ assert that
the individual object $x$ is in a given state or that $x$ has a given
property in a given $\mu $\textit{-}context, but not that $x$ has a given
property without reference to contexts, as in the standard language of QM.
Then we introduce a classical notion of probability on the set of all
sentences of $L(x)$ in Section 4 and a family of classical probability
measures on sets of $\mu $-contexts in Section 5, each element of the family
corresponding to a \textit{measurement procedure} that determines a \textit{%
macroscopic measurement context}. We can thus define a notion of \textit{%
compatibility} on the set $\mathcal{E}$ of all properties, hence a notion of 
\textit{testability} on the set of all sentences of $L(x)$, and use the
foregoing probabilities conjointly to define the notion of \textit{mean
conditional probability} on the subset of all testable sentences of $L(x)$
and the related notion of \textit{mean probability measurement}. The former
admits an interpretation that is \textit{epistemic} (in a broad sense, i.e.,
relating to our degree of knowledge/lack of knowledge), even if it is not
bound to satisfy Kolmogorov's axioms because it is obtained by averaging
over classical probability measures.

Based on the definitions and results expounded above, we focus in Section 6
on the set $\mathcal{E}$\ of all properties, on which mean conditional
probabilities induce a preorder relation $\prec $. We show that, if suitable
structural conditions are satisfied, a family of mean conditional
probabilities can be introduced, parametrized by the set $\mathcal{S}$\ of
all states, each element of which is a \textit{generalized probability
measure} on $(\mathcal{E,}\prec )$. Moreover these measures allow the
definition of a new kind of conditioning referring to a sequence of
measurement procedures that is conceptually different from classical
conditioning.

The formal scheme described above characterizes a broad class $\mathcal{T}$
of theories. Then we assume in Section 7 that QM belongs to $\mathcal{T}$,
so that states and properties can be interpreted as quantum states and
quantum properties, respectively, and the quantum probability measures
associated with states can be considered as the specific form that the
generalized probability measures defined on $\mathcal{E}$\ take in QM. Hence
we attain the following results.

(i) The nonclassical character of quantum probability can be explained in
classical terms by taking into account $\mu $-contexts. It follows in
particular that quantum probability can be given an \textit{epistemic}
rather than an \textit{ontic }interpretation in our approach.\footnote{%
We stress that our general framework does not constitutes a hidden variables
theory for QM in a standard sense. Indeed, $\mu $-contexts are associated
(generally many-to-one) with measurement procedures, not with properties or
states of the measured entity. Our perspective complies instead with Aerts'
(1986) hidden measurements approach.}

(ii) The quantum relation of compatibility on the set of properties can be
considered as the specific form that the relation of compatibility
introduced in the general framework takes in QM.

(iii) The conditional probability usually introduced in QM can be considered
as the specific form that the new kind of conditioning introduced in the
general framework takes in QM.

We conclude our treatment by observing in Section 8 that the general notions
of mean conditional probability and mean probability measurement are
conceptually close to the notions of \textit{universal average} and \textit{%
universal measurement}, respectively, introduced by Aerts and Sassoli de
Bianchi (2014, 2017). Hence our approach provides a description of
measurements of probabilities that is similar to the proposal of these
authors, which they maintain to supply a possible solution of the hoary
quantum measurement problem. We however do not make such a claim in the case
of our approach, because we supply our definition of mean probability
measurement resting on the standard notion of measurement in QM, without
entering the problematic aspects of this notion (as state reduction and
nonlocality) which arise when QM is assumed to refer to individual objects
and their properties. Nevertheless the results expounded above are
sufficient in our opinion to justify our proposal.

To close this section, let us point out an essential difference between our
approach and Khrennikov's. This author considers contexts `as a
generalization of a widely used notion of \textit{preparation procedure}'
(2009b). As we have seen, we introduce instead measurement procedures
determining macroscopic measurement contexts, each of which is associated
with a set of microscopic contexts. The latter play an essential role in our
framework, as they allow us to obtain the results resumed above, and do not
occur in Khrennikov's approach.

\section{Some remarks on QM}

As other advanced scientific theories, QM is expressed by means of a
fragment of the natural language enriched with technical terms (\textit{the
language of QM}) and is characterized by a pair $(F,I)$, with $F$ a logical
and mathematical formalism and $I$ an \textit{empirical interpretation}
which establishes connections between $F$ and an empirical domain. This
interpretation generally is \textit{indirect}, in the sense that there are
theoretical entities that are connected with the empirical domain only via 
\textit{derived} theoretical entities, and \textit{incomplete}, in the sense
that only limited ranges of values of the theoretical entities are actually
interpreted.\footnote{%
More generally, according to the \textit{standard epistemological conception}%
, or \textit{received view} (see, e.g., Braithwaite, 1953; Hempel, 1965;
Carnap, 1966), a fully-developed physical theory $T$, as QM, is expressed by
means of a metalanguage in which a \textit{theoretical language} $L_{T}$, an
observational language $L_{O}$ and \textit{correspondence rules} $R_{C}$
connecting $L_{T}$ and $L_{O}$ can be distinguished. The theoretical
apparatus of $T$, expressed by means of $L_{T}$, includes a \textit{%
mathematical structure} and, usually, an \textit{intended interpretation}
that is a direct and complete physical model of the mathematical structure.
The observational language $L_{O}$\ describes instead an empirical domain,
hence it has a semantic interpretation, so that the correspondence rules $%
R_{C}$\ provide an empirical interpretation of the mathematical structure
that is indirect and incomplete in the sense specified above.
\par
The received view has been criticized by several authors (see, e.g. Kuhn,
1962; Feyerabend, 1975) and is nowadays maintained to be outdated by some
scholars. However, we retain here some of its basic ideas that we consider
epistemologically relevant.}

To attain the results summarized in Section 1, we need to formalize an
elementary sublanguage of the language of QM. Let us therefore preliminarily
discuss some features of this language and some intuitive ideas on its
interpretation, referring to a conception of QM according to which QM deals
with individual objects and their properties, as we have anticipated in
Section 1. For the sake of simplicity we avoid distinguishing everywhere in
the following the theoretical entities from the empirical entities that
correspond to them via $I$.

First of all we recall that in most presentations of QM the notions of 
\textit{physical system}, or \textit{entity}, \textit{(physical) property}
and \textit{(physical) state} are considered as basic. Moreover, according
to some known approaches to the foundations of QM (see, e.g., Beltrametti
and Cassinelli, 1981; Ludwig, 1983) states are considered as classes of
probabilistically equivalent \textit{preparation procedures}, or \textit{%
preparing devices}, and properties as classes of probabilistically
equivalent \textit{dichotomic (yes-no) registering devices}.

A preparation procedure $\pi $ in the class $S$, when activated, produces an
individual object $x$ (which can be identified with the act of activation
itself if one wants to avoid ontological commitments). Hence, after the act
of activation, a sentence that states that $x$ is in the state $S$ is 
\textit{true} and a sentence that states that $x$ is in a state $T\neq S$ is 
\textit{false}.

Given an individual object $x$ in the state $S$, however, activating a
registering device $r$ in the class $E$ does not test whether the property $%
E $ is possessed or not by $x$ independently of the simultaneous activation
of other devices. Indeed it follows from some known proofs of the Bell and
Kochen-Specker theorems mentioned in Section 1 (see, e.g., Greenberger et
al., 1990; Mermin 1993) that, if the laws of QM have to be preserved in
every conceivable physical situation, the outcome that is obtained depends
on the set of the registering devices that are activated together with $r$,
i.e., on the macroscopic context $\boldsymbol{C}_{M}$\ determined by the
whole quantum measurement $M$ that is performed (of course, these
registering devices must be \textit{compatible}, i.e., they must belong to
different but compatible properties). Hence, one must admit that, generally,
a truth value can be assigned to a sentence which states that $x$ possesses
the property $E$ only if also a macroscopic context is specified (\textit{%
contextuality} of QM).\footnote{%
We have emphasized in some previous papers (see, e.g., Garola, 1999; Garola
and Pykacz, 2004; Garola and Sozzo, 2010; Garola and Persano, 2014) that the
epistemological clause "the laws of QM have to be preserved in every
conceivable physical situation" is essential in the proofs of Bell's and
Kochen-Specker's theorems. Nevertheless, this clause generally is not
explicitly noticed or stated, possibly because it seems to be unquestionably
justified by the outstanding success of QM. Yet it must be observed that all
the proofs mentioned above proceed ab absurdo, hypothesizing physical
situations in which noncompatible physical properties are assumed to be
simultaneously possessed by an individual object. In such situations the
quantum laws that are applied cannot be simultaneously tested, hence the
assumption that they hold anyway seems more consistent with a classical than
with a quantum view. One can therefore try to give up the aforesaid clause,
but then the proofs of Bell's an Kochen-Specker's theorems cannot be
completed. This conclusion opens the way to the attempt at recovering
noncontextual interpretations of QM (Garola, Sozzo and Wu, 2016). The
arguments in this paper, however, apply to every theory in which contexts
can be defined, irrespective of whether the results of measurements are
context-depending (locally, or also at a distance) or not.}

We observe now that, generally, the macroscopic context $\boldsymbol{C}_{M}$
determined by $M$ may be produced by many different microscopic physical
situations that cannot be distinguished at a macroscopic level (though they
can be described, in principle, by QM itself). Hence we can associate $%
\boldsymbol{C}_{M}$ with a set $\mathfrak{C}_{M}$ of \textit{microscopic
contexts} ($\mu $\textit{-contexts}; of course, $\mathfrak{C}_{M}$ could
reduce to a singleton in special cases). It is then natural to think that
the truth value of a sentence asserting that $x$ possesses the property $E$
generally depends on the $\mu $-context that is realized when $M$ is
performed. But we cannot know this $\mu $-context, hence only a probability
of it, which expresses our degree of ignorance, can be given (we naively
argue here as though the set $\mathfrak{C}_{M}$ were discrete, to avoid
technical complications).

Summing up, our analysis leads us to conclude that a truth value can be
assumed to exist consistently with QM only in the case of a sentence
asserting that an individual object $x$ possesses a property $E$ in a given $%
\mu $-context $C$, not in the case of a sentence simply asserting that $x$
possesses a property $E$. Moreover, in general this value cannot be deduced
from the laws of QM, which are probabilistic laws that make no explicit
reference to contexts.

The conclusions above have an important consequence. Every quantum
prediction concerns probabilities, hence testing it requires evaluating
frequencies of outcomes. In our present perspective, a typical test of this
kind consists in preparing a broad set of individual objects in a given
state $S$ and then performing on each of them the same quantum measurement $%
M $, which requires activating one or more (compatible) registering devices
simultaneously. The macroscopic context $\boldsymbol{C}_{M}$ then is the
same for every individual object, but the $\mu $-context $C\in \mathfrak{C}%
_{M}$ generally changes in an unpredictable way. Thus we meet two distinct
sources of randomness. The first is the state $S$ (be it a pure state or a
mixture) that may not determine univocally the properties of an individual
object in QM, even if the $\mu $-context $C$\ is given. The second is the
unpredictable change of the $\mu $-context that occurs when performing $M$
on different individual objects. We are therefore led to think that quantum
probability takes implicitly into account both sources. We will see in the
following sections that this idea can explain the non-kolmogorovian
character of quantum probability, together with the rather surprising fact
that the values of quantum probability neither depend on $\mu $-contexts nor
on macroscopic contexts (see, e.g., Mermin, 1993).

\section{The formal language $L(x)$}

As we have anticipated in Section 1, we intend to introduce in the present
paper a general probabilistic framework that may characterize a class of
theories including QM. Of course, this will be done by bearing in mind all
the suggestions following from our remarks on QM in Section 2.

As a first step we construct in this section a formal language $L(x)$ (which
formalizes, in the case of QM, an elementary sublanguage of the language of
QM, and can be considered a part of the formalism $F$). To this end we agree
to use standard symbols in set theory and logic. In particular, $^{c}$, $%
\cap $, $\cup $, $\subset $, $\backslash $, $\emptyset $ and $%
\mathbb{P}
(\Psi )$ will denote complementation, intersection, union, inclusion,
difference, empty set and power set of the set $\Psi $, respectively.
Moreover, \textsc{N} will denote the set of natural numbers.\medskip

\noindent \textbf{Definition 3.1.} \textit{Let }$\mathcal{E}$\textit{, }$%
\mathcal{S}$ \textit{and }$\mathcal{C}$\textit{\ be disjoint sets whose
elements we call }properties\textit{, }states\textit{\ and }$\mu $\textit{-}%
contexts\textit{, respectively, and let us set}

\begin{center}
$\mathcal{E}_{\mathcal{C}}=\left\{ E_{C}=(E,C)\mid E\in \mathcal{E},C\in 
\mathcal{C}\right\} $.
\end{center}

\textit{Then, we denote by }$L(x)$\textit{\ a classical predicate logic,
constructed as follows.\smallskip }

Syntax.

\textit{(i) An individual variable }$x$\textit{.}

\textit{(ii) A set }$\Pi =\mathcal{E}_{C}\cup \mathcal{S}$ \textit{of
monadic predicates.}

\textit{(iii) Connectives }$\lnot $ \textit{(not), }$\wedge $\textit{\
(and), }$\vee $\textit{\ (or).}

\textit{(iv) Parentheses }$($\textit{,}$)$\textit{.}

\textit{(v) A set }$\Psi (x)$\textit{\ of }well formed formulas\textit{\ }%
(wffs),\textit{\ obtained by applying recursively standard formation rules
in classical logic (to be precise, for every} $A\in \Pi $, $A(x)\in \Psi (x)$%
\textit{; for every} $\alpha (x)\in \Psi (x)$, $\lnot \alpha (x)\in \Psi (x)$%
\textit{; for every }$\alpha (x),\beta (x)\in \Psi (x)$, $\alpha (x)\wedge
\beta (x)\in \Psi (x)$\textit{\ and }$\alpha (x)\vee \beta (x)\in \Psi (x)$%
\textit{)}\smallskip

Semantics.

\textit{(i) A universe }$U$,\textit{\ whose elements we call }individual
objects\textit{.}

\textit{(ii) An injective mapping}

\begin{center}
$ext:A\in \Pi \longrightarrow ext(A)\in 
\mathbb{P}
(U)$.
\end{center}

\textit{(iii) The boolean sublattice }$\Theta =<ext(\Pi )>=(ext(\Pi
),^{c},\cap ,\cup )$\textit{\ of }$%
\mathbb{P}
(U)$\textit{\ generated by }$ext(\Pi )$.

\textit{(iv) The surjective mapping (still called ext by abuse of language)}

\begin{center}
$ext:\alpha (x)\in \Psi (x)\longrightarrow ext(\alpha (x))\in \Theta $
\end{center}

\noindent \textit{recursively defined by the following rules:}

\noindent \textit{for every }$A\in \Pi $\textit{, }$ext(A(x))=ext(A)$;

\noindent \textit{for every }$\alpha (x)\in \Psi (x)$\textit{, }$ext(\lnot
\alpha (x))=U\setminus ext(\alpha (x))=(ext(\alpha (x)))^{c}$\textit{;}

\noindent \textit{for every }$\alpha (x),\beta (x)\in \Psi (x)$,\textit{\ }$%
ext(\alpha (x)\wedge \beta (x))=ext(\alpha (x))\cap ext(\beta (x))$\textit{\
and }$ext(\alpha (x)\vee \beta (x))=ext(\alpha (x))\cup ext(\beta (x))$%
\textit{.}

\textit{(v) A set }$\Sigma $\textit{\ of }interpretations of the variable%
\textit{\ }$x$\textit{\ such that, for every }$\sigma \in \Sigma $\textit{,}

\begin{center}
$\sigma :x\in \left\{ x\right\} \longrightarrow \sigma (x)\in U$.
\end{center}

\textit{(vi) For every }$\sigma \in \Sigma $\textit{, a truth assignment}

\begin{center}
$\nu _{\sigma }:\alpha (x)\in \Psi (x)\longrightarrow \nu _{\sigma }(\alpha
(x))\in \left\{ t,f\right\} $
\end{center}

\noindent (\textit{where }$t$\textit{\ stands for }true\textit{\ and }$f$%
\textit{\ for }false\textit{), such that }$\nu _{\sigma }(\alpha (x))=t$%
\textit{\ iff }$\sigma (x)\in ext(\alpha (x))$\textit{\ (hence }$\nu
_{\sigma }(\alpha (x))=f$\textit{\ iff }$\sigma (x)\in (ext(\alpha
(x)))^{c}).\medskip $

The logical preorder and the Lindenbaum-Tarski algebra of $L(x)$ can then be
introduced in a standard way, as follows.\medskip

\noindent \textbf{Definition 3.2.} \textit{We denote by }$<$\textit{\ and }$%
\equiv $\textit{\ the (reflexive and transitive) relation of logical
preorder and the relation of logical equivalence on }$\Psi (x)$\textit{,
respectively, defined by standard rules in classical logic (to be precise,
for every }$\alpha (x),\beta (x)\in \Psi (x)$,\textit{\ }$\alpha (x)<\beta
(x)$\textit{\ iff, for every }$\sigma \in \Sigma $\textit{, }$\nu _{\sigma
}(\beta (x))=t$\textit{\ whenever }$\nu _{\sigma }(\alpha (x))=t$\textit{,
and }$\alpha (x)\equiv \beta (x)$\textit{\ iff }$\alpha (x)<\beta (x)$%
\textit{\ and }$\beta (x)<\alpha (x)$\textit{). Moreover we put }$\Psi
^{\prime }(x)=\Psi (x)/\equiv $\textit{\ and denote by }$<^{\prime }$ 
\textit{the partial order canonically induced by }$<$\textit{\ on }$\Psi
^{\prime }(x)$\textit{. Then }$(\Psi ^{\prime }(x),<^{\prime })$\textit{\ is
a boolean lattice (the Lindenbaum-Tarski algebra of }$L(x)$\textit{) whose
operations }$\lnot ^{\prime }$\textit{, }$\wedge ^{\prime }$\textit{,}$\vee
^{\prime }$\textit{\ are canonically induced on }$\Psi ^{\prime }(x)$\textit{%
\ by }$\lnot $\textit{, }$\wedge $\textit{, }$\vee $\textit{,
respectively).\medskip }

As stated in Definition 3.1, the language $L(x)$ is a classical predicate
logic. It has, however, some innovative features from the point of view of
the interpretation $I$. Indeed the words "states", "properties", "$\mu $%
-contexts" and "individual objects" occur in Definition 3.1 just as nouns of
elements of sets, but obviously refer to an interpretation that makes these
elements correspond to empirical notions denoted by the same nouns. Then,
each state $S$ is classified in $L(x)$ as a predicate, and an elementary wff
of the form $S(x)$ (interpreted as "the individual object $x$ is in the
state $S$")\ is argument of truth assignments, at variance with widespread
views that consider states as \textit{possible worlds} of a Kripkean
semantics in QL (see, e.g., Dalla Chiara et al., 2004). Furthermore
properties are not classified as predicates of $L(x)$. Rather, a predicate
either is a state or it is a pair $E_{C}=(E,C)$ (an elementary wff of the
form $E_{C}(x)$ is then interpreted as "the individual object $x$ has the
property $E$ in the context $C$").

\section{A contextual probability structure on $L(x)$}

We state now an assumption that is suggested by our introduction of new
entities ($\mu $-contexts) which do not occur explicitly in the formal
apparatus of QM.\medskip

\noindent \textbf{Axiom P}.\textit{\ A mapping }$\xi :ext(\alpha (x))\in
\Theta \longrightarrow \xi (ext(\alpha (x)))\in \lbrack 0,1]$\textit{\
exists such that }$\Phi =(U,\Theta ,\xi )$\textit{\ is a classical
probability space.\footnote{%
Following a standard terminology we call \textit{classical probability space 
}here any triple $(\Omega ,\Sigma ,\mu )$, where $\Omega $ is a set, $\Sigma 
$ is a Boolean $\sigma $-subalgebra of $%
\mathbb{P}
(\Omega )$, and $\mu :\Delta \in \Sigma \longrightarrow \mu (\Delta )\in
\lbrack 0,1]$ is a mapping satisfying the following conditions: (i) $\mu
(\Omega )=1$; (ii) if $\left\{ \Delta _{i}\right\} _{i\in \text{\textsc{N}}}$
is a family of pairwise disjoint elements of $\Sigma $, then $\mu (\cup
_{i}\Delta _{i})=\Sigma _{i}\mu (\Delta _{i})$.}}\medskip

Based on Axiom P we can introduce now a probability measure on $L(x)$ by
means of the following definition.\medskip

\noindent \textbf{Definition 4.1}\textit{. Let }$\Psi ^{+}(x)\subset \Psi
(x) $\textit{\ be the set of wffs of }$L(x)$\textit{\ such that, for every }$%
\beta (x)\in \Psi ^{+}(x)$,\textit{\ }$\xi (ext(\beta (x))\neq 0$\textit{,
and let }$p$\textit{\ be a binary mapping such that}

\begin{center}
$p:(\alpha (x),\beta (x))\in \Psi (x)\times \Psi ^{+}(x)\longrightarrow
p(\alpha (x)\mid \beta (x))=\frac{\xi (ext(\alpha (x))\cap ext(\beta (x)))}{%
\xi (ext(\beta (x)))}\in \lbrack 0,1].$
\end{center}

\noindent \textit{We say that the pair }$(\Phi ,p)$\textit{\ is a }$\mu $%
\textit{-}contextual probability structure\textit{\ on }$L(x)$\textit{\ and
that }$p(\alpha (x)\mid \beta (x))$\textit{\ is the }$\mu $\textit{-}%
contextual conditional probability\textit{\ of }$\alpha (x)\ $\textit{given }%
$\beta (x)$\textit{. Moreover, whenever }$ext(\beta (x))=U$\textit{\ we say
that }$p(\alpha (x)\mid \beta (x))$ \textit{is the }$\mu $\textit{-}%
contextual absolute probability\textit{\ of }$\alpha (x)$\textit{\ and
simply write }$p(\alpha (x))$\textit{\ in place of }$p(\alpha (x)\mid \beta
(x))$\textit{.\medskip }

The terminology introduced in Definition 4.1 (where the word $\mu $\textit{%
-contextual} underlines the dependence of probabilities on $\mu $-contexts
through the wffs of $L(x)$)\textit{, }is justified by the following
statement.\medskip

\noindent \textbf{Proposition 4.1.} \textit{Let }$\beta (x)\in \Psi ^{+}(x)$%
\textit{. Then, the mapping}

\begin{center}
$p_{\beta (x)}:\alpha (x)\in \Psi (x)\longrightarrow p(\alpha (x)\mid \beta
(x))\in \lbrack 0,1]$
\end{center}

\noindent \textit{satisfies the following conditions.}

\textit{(i) Let }$\alpha (x)\in \Psi (x)$\textit{\ be such that }$ext(\alpha
(x))=U$\textit{\ (equivalently, }$\alpha (x)$\textit{\ }$\equiv \alpha
(x)\vee \lnot \alpha (x)$\textit{). Then, }$p_{\beta (x)}(\alpha (x))=1$%
\textit{.}

\textit{(ii) Let }$\alpha _{1}(x),\alpha _{2}(x)\in \Psi (x)$\textit{\ be
such that }$ext(\alpha _{1}(x))\cap ext(\alpha _{2}(x))=\emptyset $\textit{\
(equivalently, }$\alpha _{1}(x)<\lnot \alpha _{2}(x)$\textit{). Then, }$%
p_{\beta (x)}(\alpha _{1}(x)\vee \alpha _{2}(x))=p_{\beta (x)}(\alpha
_{1}(x))+p_{\beta (x)}(\alpha _{2}(x))$.\smallskip

Proof. Straightforward.\medskip

Proposition 4.1 shows indeed that, for every $\beta (x)\in \Psi ^{+}(x)$%
\textit{, }$p_{\beta (x)}$ is a \textit{probability measure} on $(\Psi
(x),\lnot ,\wedge ,\vee )$.\medskip

\noindent \textbf{Examples.} Let $E,F\in \mathcal{E}$, $S,T\in \mathcal{S}$, 
$C,D\in \mathcal{C}$, and let $F_{D}(x),S(x)\in \Psi ^{+}(x)$. Then, we
obtain from Definition 4.1:\medskip

(i) $p(E_{C}(x)\mid F_{D}(x))=\frac{\xi (ext(E_{C}(x))\cap ext(F_{D}(x)))}{%
\xi (ext(F_{D}(x)))}$;\medskip

(ii) $p(E_{C}(x)\mid S(x))=\frac{\xi (ext(E_{C}(x))\cap ext(S(x)))}{\xi
(ext(S(x)))}$;\medskip

(iii) $p(T(x)\mid S(x))=\frac{\xi (ext(T(x))\cap ext(S(x)))}{\xi (ext(S(x)))}
$.\medskip

Example (iii) is especially interesting because it shows that the $\mu $%
-contextual conditional probabilities do not always depend on $\mu $%
-contexts.

By using Axiom P we have thus introduced $\mu $-contextual conditional and
absolute probabilities on $L(x)$. We stress that the $\mu $-contextual
probability structure introduced in Definition 4.1 is basically classical,
hence these probabilities admit an epistemic interpretation. In other words,
they can be considered as indexes of our lack of knowledge of the truth
assignments on $L(x)$.

\section{Measurements and mean probabilities}

Based on the notions introduced in Sections 3 and 4, we intend to supply in
this section a theoretical description of measurements testing
probabilities. To this end, let us observe that our remarks in Section 2
suggest that a test of the probability of a wff $\alpha (x)\in \Psi (x)$%
\textit{\ }consists in choosing a measurement that checks all the properties
that occur in $\alpha (x)$ (hence these properties must be compatible) on an
individual object, performing it on a large number of individual objects,
and then evaluating the frequencies of the outcomes that have been obtained.
Moreover, the theoretical description of this test must refer to a
probability measure defined on some set of $\mu $-contexts, to take into
account our limited knowledge of the $\mu $-context that must be associated
with each implementation of the measurement on an individual object. Bearing
in mind these requirements, we introduce the following assumption.\medskip

\noindent \textbf{Axiom M}. \textit{Every }$E\in \mathcal{E}$\textit{\ is
associated with a set }$\mathcal{M}_{E}$\textit{\ of }measurement procedures%
\textit{,\footnote{%
We denote here abstract measurement procedures with the same symbols that we
have used in Section 2 to denote quantum measurements. We shall see in
Section 7 that the latter can be considered as the specific form that the
former take in QM.} and every }$M\in \mathcal{M}_{E}$\textit{\ determines a }%
macroscopic measurement context\textit{\ C}$_{M}$\textit{\ associated with a
classical probability space }$(\mathcal{C}_{M},\Sigma _{M},\nu _{M})$\textit{%
, where }$\mathcal{C}_{M}$\textit{\ is a set of }$\mu $\textit{-contexts
and, for every }$C\in \mathcal{C}_{M}$\textit{, }$\left\{ C\right\} $ 
\textit{belongs to }$\Sigma _{M}$\textit{.\medskip }

We have seen in Section 2 that a quantum measurement may require that more
than one property be simultaneously tested. We are thus naturally led to
introduce the notions of \textit{compatibility}, \textit{testability} and 
\textit{conjoint testability} in our present framework, as follows.\medskip

\noindent \textbf{Definition 5.1.} \textit{Let }$\left\{ E,F,...\right\} $%
\textit{\ be a countable set\ of properties of }$L(x)$\textit{. We say that }%
$E,F,...$\textit{\ are }compatible\textit{\ iff }$\mathcal{M}_{E}\cap 
\mathcal{M}_{F}\cap ...\neq \emptyset $\textit{, and denote by }$k$\textit{\
the }compatibility relation\textit{\ on }$\mathcal{E}$\textit{\ defined by
setting}

\begin{center}
\textit{\ for every }$E,F\in \mathcal{E}$\textit{, }$EkF$\textit{\ }iff$\ E$%
\textit{\ and }$F$ \textit{are compatible.}
\end{center}

\textit{Moreover, let }$\alpha (x)\in \Psi (x)$\textit{\ and let }$E,F,...$%
\textit{\ be the properties that occur in the formal expression of }$\alpha
(x)$\textit{\ (with indexes in }$\mathcal{C}$\textit{). Then we say that} $%
\alpha (x)$ \textit{is }testable\textit{\ iff the following conditions hold.}

\textit{(i) }$E,F,...$\textit{\ are compatible.}

\textit{(ii) }$E,F,...$\textit{\ occur in the formal expression of }$\alpha
(x)$\textit{\ with the same index }$C$\textit{\ and a macroscopic
measurement procedure }$M\in \mathcal{M}_{E}\cap \mathcal{M}_{F}\cap ...$ 
\textit{exists such that }$C\in \mathcal{C}_{M}$.

\textit{Finally, let }$\left\{ \alpha (x),\beta (x),...\right\} $\textit{\
be a countable set of wffs of }$\Psi (x)$,\textit{\ We say that }$\alpha
(x),\beta (x),...$\textit{\ are }jointly testable\textit{\ iff the wff }$%
\alpha (x)\wedge \beta (x)\wedge ...$\textit{\ is testable. Then we denote
by }$\Psi ^{T}(x)$ \textit{the set of all testable propositions of }$\Psi
(x) $ \textit{and, for every }$\alpha (x)\in \Psi ^{T}(x)$,\textit{\ we
write }$\alpha _{M}^{C}(x)$\textit{\ in} \textit{place of }$\alpha (x)$%
\textit{\ whenever explicit reference to the measurement procedure }$M$%
\textit{\ and to the }$\mu $\textit{-context }$C$\textit{\ defined in (ii)
must be done.\medskip }

We can now state the following proposition.\medskip

\noindent \textbf{Proposition 5.1.}\textit{\ (i) The binary relation }$k$%
\textit{\ on }$\mathcal{E}$\textit{\ introduced in Definition 5.1 is
reflexive and symmetric, but, generally, not transitive.}

\textit{(ii) Let }$E\in \mathcal{E}$\textit{, }$M\in M_{E}$\textit{\ and }$%
C\in \mathcal{C}_{M}$\textit{. Then, }$E_{C}(x)\in \Psi ^{T}(x)$\textit{.}

\textit{(iii) Let }$M$\textit{\ be a measurement procedure, }$C,C^{\prime
}\in M$\textit{\ and }$C^{\prime }\neq C$\textit{.} \textit{Moreover, for
every }$\alpha _{M}^{C}(x)\in \Psi ^{T}(x)$, \textit{let }$\alpha
_{M}^{C^{\prime }}(x)$\textit{\ be the wff obtained from }$\alpha
_{M}^{C}(x) $\textit{\ by replacing }$C$ \textit{with }$C^{\prime }$. 
\textit{Then, }$\alpha _{M}^{C^{\prime }}(x)\in \Psi ^{T}(x)$.$\smallskip $

Proof. Straightforward.\textit{\medskip }

Of course, in every theory of the class that we are considering, each
measurement procedure $M$ provides a theoretical description, via an
empirical interpretation $I$ (see Section 2) of a concrete measurement.
Then, it remains to understand what one actually tests when evaluating the
frequencies of outcomes obtained as explained above. It is apparent indeed
that such a test does not refer to the $\mu $-contextual conditional
probabilities introduced in Definition 4.1, because we cannot know nor fix
the $\mu $-context associated with each implementation of the measurement
(hence $\mu $-contextual probabilities must be classified as theoretical
entities that can be interpreted only indirectly, see Section 2). But the
unpredictable change of $\mu $-context that generally occurs when performing
the measurement on different individual objects suggests that one actually
tests a mean of contextual $\mu $-conditional probabilities over the family $%
\left\{ \alpha _{M}^{C}(x)\right\} _{C\in \mathcal{C}_{M}}$. The following
definition and assumption formalize this idea.\medskip

\noindent \textbf{Definition 5.2.} \textit{Let }$\alpha (x),\beta (x)\in
\Psi ^{T}(x)$\ \textit{\ be jointly testable and let }$E,F,...\in \mathcal{E}
$\textit{\ be the properties that occur in one or both the formal
expressions of }$\alpha (x)$\textit{\ and }$\beta (x)$\textit{. Furthermore,
let }$\beta (x)\in \Psi ^{+}(x)$\textit{. For every }$M\in \mathcal{M}%
_{E}\cap \mathcal{M}_{F}\cap ...$\textit{\ we put}

\begin{center}
$<p(\alpha _{M}^{C}(x)\mid \beta _{M}^{C}(x))>_{\mathcal{C}_{M}}=\sum_{C\in 
\mathcal{C}_{M}}\nu _{M}(\{C\})p(\alpha _{M}^{C}(x)\mid \beta _{M}^{C}(x))$.
\end{center}

\textit{Moreover, whenever the\ following\ equality\ holds\ for every }$%
M,N\in \mathcal{M}_{E}\cap \mathcal{M}_{F}\cap ...$

\begin{center}
$<p(\alpha _{M}^{C}(x)\mid \beta _{M}^{C}(x))>_{\mathcal{C}_{M}}=<p(\alpha
_{N}^{C}(x)\mid \beta _{N}^{C}(x))>_{\mathcal{C}_{N}}$,
\end{center}

\noindent \textit{we omit the symbols} $M$, $N$, $C$, $D$, $\mathcal{C}_{M}$%
\textit{\ and }$\mathcal{C}_{N}$\textit{, and say that }$<p(\alpha (x)\mid
\beta (x))>$ \textit{is the }mean conditional probability\textit{\ of }$%
\alpha (x)$\textit{\ given }$\beta (x)$.\medskip

Based on Definition 5.2 we maintain in the following that performing the
measurement corresponding (via $I$) to a measurement procedure $M\in 
\mathcal{M}_{E}\cap \mathcal{M}_{F}\cap ...$ on a large number of individual
objects provides a test of $<p(\alpha (x)\mid \beta (x))>$\medskip , or,
briefly, a \textit{mean probability measurement}.\medskip

\noindent \textbf{Axiom C}. \textit{Mean conditional probability (hence mean
probability measurements) do exist for every pair }$(\alpha (x),\beta (x))$%
\textit{\ of jointly testable wffs such that }$\beta (x)\in \Psi ^{+}(x)$%
\textit{.\medskip }

It follows from Definition 5.2 and Axiom C that mean conditional
probabilities take into account two different kinds of ignorance. First, the
lack of knowledge about the truth assignments on $L(x)$ mentioned at the end
of Section 4. Second, the ignorance of the $\mu $-context to be associated
with a measurement when this measurement is performed. Hence mean
conditional probabilities admit an epistemic interpretation even if they are
not bound to satisfy Kolmogorov's axioms, for they are average quantities.

To close this section, let us observe that our present perspective is
supported by some previous research in the literature. Indeed, as we have
anticipated in Section 1, mean conditional probabilities and mean
probability measurements are conceptually similar to the universal averages
and the universal measurements, respectively, introduced by Aerts and
Sassoli de Bianchi (2014, 2017). Moreover, the recognition that two kinds of
lack of knowledge occur when a measurement is performed fits in well with
similar remarks of these authors.\footnote{%
We recall that the Aerts and Sassoli de Bianchi proposal finds its roots in
the \textit{hidden measurement approach} (see, e.g., Aerts, 1986). This
approach led the author to introduce \textit{state property systems} (see,
e.g., Aerts, 1999), that successively evolved in the \textit{%
state-context-property} (SCoP) formalism (see, e.g., Aerts and Gabora, 2005;
this formalism was mainly used for working out a theory of concepts, in
particular in the field of quantum cognition). It is then possible to show
that the SCoP formalism can be (partially) translated into the formalism
developed in the present paper, and conversely, which explains the
conceptual similarities pointed out above. For the sake of brevity we do not
deal with this issue in detail here.}

\section{Q-probability}

The set $\mathcal{E}$ of all properties has a relevant role in QM, hence we
focus on it in the present section.

By using the notion of mean conditional probability introduced in Section 5,
we firstly define an order structure on $\mathcal{E}$, as follows.\medskip

\noindent \textbf{Definition 6.1}. \textit{Let }$E\in \mathcal{E}$\textit{, }%
$M\in \mathcal{M}_{E}$\textit{, }$C\in \mathcal{C}_{M}$\textit{, }$S\in 
\mathcal{S}$\textit{, let }$S(x)\in \Psi ^{+}(x)$, \textit{and let}

\begin{center}
$P_{S}:E\in \mathcal{E}\longrightarrow P_{S}(E)\in \lbrack 0,1]$
\end{center}

\noindent \textit{be the mapping defined by setting}

\begin{center}
$P_{S}(E)=<p(E_{C}(x)\mid S(x))>=\sum_{C\in \mathcal{C}_{M}}\nu
_{M}(\{C\})p(E_{C}(x)\mid S(x))=\medskip $

$\sum_{C\in \mathcal{C}_{M}}\nu _{M}(\{C\})\frac{\xi (Ext(E_{C}(x))\cap
Ext(S(x)))}{\xi (Ext(S(x)))}$\textit{.}
\end{center}

\textit{Then, \ we denote by }$\prec $\textit{\ and }$\approx $\textit{\ the
preorder and the equivalence relation on }$\mathcal{E}$\textit{,
respectively, defined by setting, for every }$E,F\in \mathcal{E}$\textit{,}

\begin{center}
$E\prec F$\textit{\ }iff\textit{, for every }$S\in \mathcal{S}$\textit{, }$%
P_{S}(E)\leq P_{S}(F)$
\end{center}

\noindent \textit{and}

\begin{center}
$E\approx F$\textit{\ }iff\textit{\ }$E\prec F$\textit{\ and }$F\prec E$%
\textit{\ \medskip }
\end{center}

It is now important to consider a special case that allows us to connect our
present framwork with QM. We therefore introduce the following
definition.\medskip

\noindent \textbf{Definition 6.2}. \textit{Let }$\prec $\textit{\ be a
partial order on }$\mathcal{E}$\textit{\ and let }$(\mathcal{E},\prec )$%
\textit{\ be an orthocomplemented lattice. We denote meet, join,
orthocomplementation, least element and greatest element of }$(\mathcal{E}%
,\prec )$\ \textit{by }$\Cap $\textit{, }$\Cup $\textit{, }$^{\bot }$\textit{%
, }$\mathsf{O}$\textit{\ and }$\mathsf{U}$\textit{, respectively. Moreover,
we denote by }$\bot $\textit{\ the (binary) orthogonality relation
canonically induced by }$^{\bot }$\textit{\ on }$(\mathcal{E},\Cap \mathit{,}%
\Cup \mathit{,}^{\bot })$\textit{\footnote{%
We recall that $^{\bot }\mathcal{\ }$is a unary operation on $(\mathcal{E}%
,\prec )$ such that, for every $E,F\in \mathcal{E}$, $E^{\bot \bot }=E$, $%
E\prec F$ implies $F^{\bot }\prec E^{\bot }$, $E\Cap E^{\bot }=\emph{O}$ and 
$E\Cup E^{\bot }=\emph{I}$. Then $\bot $\ is the non-reflexive and symmetric
binary relation on $\mathcal{E}$\ defined by setting, for every $E,F\in 
\mathcal{E}$, $E\bot F$ iff $E\prec F^{\bot }.$}. Then, for every }$S\in 
\mathcal{S}$\textit{, we say that }$P_{S}$\textit{\ is a }generalized
probability measure\textit{\ on }$(\mathcal{E},\Cap \mathit{,}\Cup \mathit{,}%
^{\bot })$\ \textit{iff it satisfies the following conditions.}

\textit{(i) }$P_{S}(\mathsf{U})=1$.

\textit{(ii) If }$\left\{ E_{1},E_{2},...\right\} $ \textit{is a countable
set of properties of }$\mathcal{E}$\textit{\ and }$E_{1},E_{2},...$\textit{\
are pairwise disjoint (i.e., for every }$k$\textit{, }$l$\textit{, }$%
E_{k}\bot E_{l}$),\textit{\ then}

\begin{center}
$P_{S}(\Cup _{k}E_{k})=\sum_{k}P_{S}(E_{k})$.
\end{center}

\textit{Whenever }$P_{S}$\textit{\ is a generalized probability measure on }$%
(\mathcal{E},\Cap \mathit{,}\Cup \mathit{,}^{\bot })$, \textit{for every }$%
E\in \mathcal{E}$\textit{\ we say that }$P_{S}(E)$\textit{\ is the }%
Q-probability\textit{\ of }$E$\textit{\ given}$\mathcal{\ }S$\textit{%
.\medskip }

Definition 6.2 implies that a generalized probability measure $P_{S}$\ does
not satisfy Kolmogorov's axioms if $(\mathcal{E},\Cap \mathit{,}\Cup \mathit{%
,}^{\bot })$\ is not a boolean lattice. Nevertheless the Q-probability $%
P_{S}(E)$\ of a property $E\in \mathcal{E}$\ given $S$ admits an epistemic
interpretation and can be empirically checked, as it is a special case of
the mean conditional probability introduced in Definition 5.2. It is then
natural to wonder whether a \textit{conditional Q-probability} of a property 
$E\in \mathcal{E}$ given another property $F\in \mathcal{E}$\ can be defined
by means of $P_{S}$, generalizing standard procedures in classical
propositional logic. But if one tries to put

\begin{center}
$P_{S}(E\mid F)=\frac{P_{S}(E\Cap F)}{P_{S}(F)}$,
\end{center}

\noindent then the mapping

\begin{center}
$P_{SF}:E\in \mathcal{E}\longrightarrow P_{S}(E\mid F)\in \lbrack 0,1]$
\end{center}

\noindent is not a generalized probability measure on $(\mathcal{E},\Cap 
\mathit{,}\Cup \mathit{,}^{\bot })$\ whenever this lattice is not boolean.
Indeed, consider a property $E=E_{1}\Cup E_{2}$, with $E_{1},E_{2}\in 
\mathcal{E}$ and $E_{1}\bot E_{2}$. Then, we obtain

\begin{center}
$P_{SF}(E)=P_{SF}(E_{1}\Cup E_{2})=P_{S}(E_{1}\Cup E_{2}\mid F)=\frac{%
P_{S}((E_{1}\Cup E_{2})\Cap F)}{P_{S}(F)}$,
\end{center}

\noindent which is generally different from

\begin{center}
$\frac{P_{S}((E_{1}\Cap F)\Cup (E_{2}\Cap F))}{P_{S}(F)}=P_{S}(E_{1}\mid
F)+P_{S}(E_{2}\mid F)=P_{SF}(E_{1})+P_{SF}(E_{2})$
\end{center}

\noindent whenever $(\mathcal{E},\Cap \mathit{,}\Cup \mathit{,}^{\bot })$\
is not distributive.\medskip

To overcome this difficulty one can intuitively refer to a sequence of two
measurements and introduce a non-standard kind of conditional probability,
as follows.\medskip

\noindent \textbf{Definition 6.3}. \textit{Let }$E\in \mathcal{E}$\textit{\
and let us put }$\mathcal{S}_{E}=\left\{ S\in \mathcal{S}\mid P_{S}(E)\neq
0\right\} $\textit{. We say that a measurement procedure} $M\in \mathcal{M}%
_{E}$\ \textit{is of }first kind\textit{\ iff it is associated with a mapping%
}

\begin{center}
$t_{E}:S\in \mathcal{S}_{E}\longrightarrow t_{E}(S)\in \mathcal{S}_{E}$
\end{center}

\noindent \textit{such that }$P_{t_{E}(S)}(E)=1$.\textit{\ For every }$F\in 
\mathcal{E}$ \textit{we then put }

\begin{center}
$P_{S}(F\Vert E)=P_{t_{E}(S)}(F)$\textit{.}
\end{center}

\textit{Moreover, let }$(\mathcal{E},\prec )$\textit{\ be an
orhocomplemented lattice\ and let }$P_{S}$\textit{\ and }$P_{t_{E}(S)}$%
\textit{\ be generalized probability measures on }$(\mathcal{E},\prec )$%
\textit{. Then we say that }$P_{S}(E\Vert F)$\textit{\ is the }conditional
Q-probability\textit{\ of }$E$\textit{\ given }$F$\textit{\ and }$S$\textit{%
.\medskip }

If a measurement corresponding (via $I$) to a first kind measurement
procedure $M\in M_{E}$ exists and the conditions at the end of Definition
6.3 are fulfilled, then $P_{S}(E\Vert F)$ can be tested whenever $S\in 
\mathcal{S}_{F}$, as Axiom C implies that $P_{t_{E}(S)}(E)$ can always be
tested (but no analogous of the Bayes theorem can be stated for conditional
Q-probabilities). Definition 6.3 thus introduces a non-standard conditional
probability on $(\mathcal{E},\prec )$ that coexists with the (classical) $%
\mu $-conditional probability introduced in Definition 4.1 (which instead
cannot be tested directly and has the status of a purely theoretical notion,
as we have seen in Section 6).\medskip

\section{Back to QM}

Axiom P in Section 4 and axioms M and C in Section 5 characterize a broad
class $\mathcal{T}$ of theories, even if they have been introduced mainly by
bearing in mind QM. They do not occur in the standard formulation of QM, but
if we assume that they underlie QM, so that QM belongs to $\mathcal{T}$, we
can explain some relevant aspects of QM\ in terms of the general notions
characterizing $\mathcal{T}$ and obtain a new perspective on quantum
probability.

To attain these results let us firstly recall that in Hibert space QM the
following mathematical representation is adopted.

Entity (physical system) $\Longrightarrow $ Hilbert space $\mathcal{H}$.

State $S\in \mathcal{S}$ $\Longrightarrow $ Density operator $\rho _{S}$ on $%
\mathcal{H}$.

Property $E\in \mathcal{E}$ $\Longrightarrow $ Orthogonal projection
operator $P_{E}$ on $\mathcal{H}$.

Furthermore, the set of all orthogonal projection operators on $\mathcal{H}$
is an orthomodular lattice in which the partial order is defined
independently of any probability measure. Hence, the representation above
induces on $\mathcal{E}$\ an order, that we denote by $\ll $, and $(\mathcal{%
E,}\ll )$\ \ is an orthomodular lattice.

Secondly, let us recall that the Born rule associates a probability value $Tr%
\left[ \rho _{S}P_{E}\right] $ (that does not depend on any context) with
the pair $(E,S)$. Hence a quantum probability

\begin{center}
$Q_{S}:E\in \mathcal{E}\longrightarrow Tr\left[ \rho _{S}P_{E}\right] \in
\lbrack 0,1]$
\end{center}

\noindent is defined which is said to be a \textit{generalized probability
measure} on $(\mathcal{E,}\ll )$\ (see, e.g., Beltrametti and Cassinelli,
1981). Moreover, the family $\left\{ Q_{S}\right\} _{S\in \mathcal{S}}$ is 
\textit{ordering} on $(\mathcal{E,}\ll )$ (ibid.), which means that the
order induced by it on $\mathcal{E}$ coincides with $\ll $. Therefore the
lattice structure of $(\mathcal{E,}\ll )$ can be seen as induced by $\left\{
Q_{S}\right\} _{S\in \mathcal{S}}$.

Based on the above remarks, and assuming that QM belongs to $\mathcal{T}$,
the order $\ll $\ and the quantum probability $Q_{S}$ can be considered as
the specific forms that the order $\prec $\ and the mapping $P_{S}$\ (see
Definition 6.1), respectively, take in QM. We thus obtain

\begin{center}
$P_{S}(E)=P(E\mid S)=Q_{S}(E)=Tr\left[ \rho _{S}P_{E}\right] $.
\end{center}

If the quantum probability $Q_{S}$ replaces $P_{S}$ in the conditions (i)
and (ii) stated in Definition 6.2, then these conditions are satisfied,
which makes the above classification of $Q_{S}$\ as a generalized
probability measure consistent with Definition 6.2.

The above interpretation of quantum probability leads to consider it as a
mean conditional probability (see Definition 5.2). This explains its
non-classical character and shows that it can be considered epistemic, at
variance with its standard ontic interpretation (see Section 5). Our main
goal in this paper has thus been achieved.

Let us denote now by $\kappa $ the compatibility relation introduced in QM
on the set of all properties by setting, for every pair $(E,F)$ of
properties, $E\kappa F$ iff $\left[ P_{E},P_{F}\right] =0$. This relation is
reflexive and symmetric but not transitive. Hence it can be considered as
the specific form that the relation $k$ introduced in Definition 5.1 takes
in QM.

Coming to quantum measurements, let us remind that first kind quantum
measurements exist in QM (see e.g., Piron, 1976; Beltrametti and Cassinelli,
1981), and that the L\"{u}ders rule states that, whenever a first kind
(ideal) quantum measurement of a property $E$ is performed on an individual
object $x$ in the state $S$ and the yes outcome is obtained, then the state
of the object after the measurement is described by the density operator $%
\frac{P_{E}\rho _{S}P_{E}}{Tr\left[ \rho _{S}P_{E}\right] }$. Let us
therefore denote by $D\left( \mathcal{H}\right) $ the set of all density
operators on $\mathcal{H}$. Then the mapping

\begin{center}
$\tau _{E}:\rho _{S}\in D\left( \mathcal{H}\right) \longrightarrow \tau
_{E}(\rho _{S})=\frac{P_{E}\rho _{S}P_{E}}{Tr\left[ \rho _{S}P_{E}\right] }%
\in D\left( \mathcal{H}\right) $
\end{center}

\noindent can be considered as the specific form that the mapping $t_{E}$
introduced in Definition 6.3 takes in QM.

Finally, we recall that the conditional probability $Q_{S}(E\mid F)$, in a
state $S$, of a property $E$ given a property $F$, is defined in QM by
referring to a quantum measurement of $E$ after a quantum measurement of $F$
on an individual object in the state $S$, and it is given by $\frac{Tr\left[
P_{E}P_{F}\rho _{S}P_{F}P_{E}\right] }{Tr\left[ P_{E}\rho _{S}P_{E}\right] }$%
. Hence this quantity can be considered as the specific form that the
conditional Q-probability of $E$ given $F$ and $S$ introduced in Definition
6.3 takes in QM. We thus obtain

\begin{center}
$P_{S}(E\Vert F)=Q_{S}(E\mid F)=\frac{Tr\left[ P_{E}P_{F}\rho _{S}P_{F}P_{E}%
\right] }{Tr\left[ P_{E}\rho _{S}P_{E}\right] }$.
\end{center}

\section{Closing remarks}

As we have observed in Sections 1 and 5, our mean conditional probabilities
and mean probability measurements are conceptually similar to the universal
averages and universal measurements, respectively, introduced by Aerts and
Sassoli de Bianchi (2014, 2017). In particular, our recognition that mean
conditional probability summarizes two kinds of lack of knowledge fits in
well with the perspective of these authors. However, Aerts and Sassoli de
Bianchi uphold that their proposal leads to a possible solution of the
quantum measurement problem. Our approach, instead, has been conceived to
show that nonclassical (yet epistemic) probabilities may occur as a
consequence of contextuality in a broad class of theories. By assuming that
QM belongs to this class we obtain an explanation of some typical features
of QM in terms of more primitive notions. In particular, the compatibility
relation on the set of all physical properties and the quantum notion of
conditional probability can be seen as special cases of general notions that
can be introduced whenever the links between contextuality and nonclassical
probability are recognized. More important, we obtain an epistemic
interpretation of quantum probability, notwithstanding its nonclassical
structure, that opposes its standard ontic interpretation. We cannot provide
instead an explanation of the reduction of the state vector carried out by a
quantum measurement in our framework, or avoid the "paradox" of nonlocality
of QM (see Section 1 and footnote 3).\bigskip

\bigskip

\textbf{Acknowledgement}. We thank Dr. Karin Verelst for useful discussions
at the Symposium "Worlds of Entanglement" in Brussels.\bigskip

\bigskip

\bigskip

\begin{center}
\textbf{BIBLIOGRAPHY}

\medskip
\end{center}

Aerts, D. (1986). A possible explanation for the probabilities of quantum
mechanics. \textit{J. Math. Physics} \textbf{27}, 202-210.

Aerts, D. (1999). Foundations of quantum physics: a general realistic and
operational approach. \textit{Int. J. Theor. Phys.} \textbf{38}, 289-358.

Aerts, D. and Gabora, L. (2005). A state-context-property model of concepts
and their combinations ii: A Hilbert space representation. \textit{Kibernetes%
} \textbf{34}, 176-204.

Aerts, D. and Sassoli de Bianchi, M. (2014). The extended Bloch
representation of quantum mechanics and the hidden-measurement solution of
the measurement problem. \textit{Ann. Phys}. \textbf{351}, 975-1025.

Aerts, D. and Sassoli de Bianchi, M. (2017). \textit{Universal Measurements.
How to Free Three Birds in One Move.} World Scientific, Singapore.

Bell, J.S. (1964). On the Einstein-Podolski-Rosen Paradox. \textit{Physics} 
\textbf{1}, 195--200.

Bell, J.S. (1966). On the Problem of Hidden Variables in Quantum Mechanics. 
\textit{Rev. Mod. Phys.} \textbf{38}, 447--452.

Beltrametti, E. and Cassinelli, G. (1981). \textit{The Logic of Quantum
Mechanics.} Reading (MA), Addison-Wesley.

Birkhoff, G. and von Neumann, J. (1936). The Logic of Quantum Mechanics. 
\textit{Ann. Math.} \textbf{37}, 823--843.

Braithwaite, R.B. (1953). \textit{Scientific Explanation}. Cambridge
University Press, Cambridge.

Busch, P., Lahti, P.J. and Mittelstaedt, P. (1996). \textit{The Quantum
Theory of Measurement}. Springer, Berlin.

Carnap, R. (1966). \textit{Philosophical Foundations of Physics}. Basic
Books Inc., New York.

Dalla Chiara, M. L., Giuntini, R. and Greechie, R. (2004). \textit{Reasoning
in Quantum Theory}. Kluwer, Dordrecht.

Feyerabend, F. (1975). \textit{Against Method: Outline of an Anarchist
Theory of Knowledge}. New Left Books, London.

Garola, C. (1999). Semantic realism: a new philosophy for quantum physics. 
\textit{Int. J. Theor. Phys.} \textbf{38}, 3241-3252.

Garola, C. and Pykacz, J. (2004). Locality and measurement within the SR
model for an objective interpretation of quantum mechanics. \textit{Found.
Phys.} \textbf{34}, 449-475.

Garola, C. and Persano, M. (2014). Embedding quantum mechanics into a
broader noncontextual theory. \textit{Found. Sci.} \textbf{19}, 217-239.

Garola, C. and Sozzo, S. (2010). Realistic aspects in the standard
interpretation of quantum mechanics. \textit{Humana.ment. J. Phil. Stud.} 
\textbf{13}, 81-101.

Garola, C. and Sozzo, S. (2013). Recovering quantum logic within an extended
classical framework. \textit{Erkenntnis} \textbf{78}, 399-314.

Garola, C., Sozzo, S. and Wu, J. (2016). Outline of a generalization and a
reinterpretation of quantum mechanics recovering objectivity. \textit{Int.
J. Theor. Phys.} \textbf{55}, 2500-2528.

Greenberger, D.M., Horne, M.A., Shimony, A. and Zeilinger, A. (1990). Bell's
theorem without inequalities. \textit{Am. J. Phys.} \textbf{58}, 1131-1143.

Hempel, C. C. (1965). \textit{Aspects of Scientific Explanation}. Free
Press, New York.

Kochen, S. and Specker, E. P. (1967). The Problem of Hidden Variables in
Quantum Mechanics. \textit{J. Math. Mech.} \textbf{17}, 59--87.

Khrennikov, A. (2009). \textit{Contextual approach to quantum formalism}.
Springer, New York.

Kuhn, T.S. (1962). \textit{The Structure of Scientific Revolution}. Chicago
University Press, Chicago.

Ludwig, G. (1983). \textit{Foundations of Quantum Mechanics I}. Springer,
New York.

Mermin, N.D. (1993). Hidden variables and the two theorems of John Bell. 
\textit{Rev. Mod. Phys.} \textbf{65}, 803-815.

Piron, C. (1976). \textit{Foundations of Quantum Physics}. Benjamin, Reading
(MA).

\end{document}